\documentclass{aa}
\usepackage{graphicx}
\begin{document}
   \title{A catalog of warps in spiral and lenticular galaxies in the Southern hemisphere}

   \author{M. L. S\'anchez-Saavedra \inst{1}, E. Battaner \inst{1}, A. Guijarro \inst{2}, M. L\'opez-Corredoira \inst{3} \and N. Castro-Rodr\'{\i}guez. \inst{3,4}}

\institute{Dpto. F\'{\i}sica Te\'orica y del Cosmos, Universidad de Granada, Avd. Fuentenueva, E-18002, Granada, Spain
          \and 
          Centro Astron\'omico Hispano Alem\'an, E-04080, Almer\'{\i}a, Spain
          \and
          Astronomisches Institut der Universitat Basel, Venusstrasse 7, Binningen, Switzerland
          \and 
          Instituto de Astrof\'{\i}sica de Canarias, 38205 La Laguna, Tenerife, Spain}		
\date{ }
\authorrunning{M. L. S\'anchez-Saavedra et al.}
\titlerunning{A catalog of warps in spiral and lenticular galaxies}

   \offprints{M. L. S\'anchez-Saavedra, malusa@ugr.es}

   \abstract{ A catalog of optical warps of galaxies is
 presented. This can be considered complementary to that reported by
 S\'anchez-Saavedra et al. (\cite{sanchez-saavedra}), with 42 galaxies
 in the northern hemisphere, and to that by Reshetnikov \& Combes
 (\cite{reshetnikov99}), with 60 optical warps. 
 The limits of the present catalog are: logr25 $>$ 0.60, $B_{t}$ $<$ 14.5,
 $\delta$(2000) $<$ 0$^o$, -2.5 $<$ t $<$ 7. Therefore, lenticular
 galaxies have also been considered. This catalog lists 150 warped
 galaxies out of a sample of 276 edge-on galaxies and covers the whole
 southern hemisphere, except the Avoidance Zone. 
 It is therefore very suitable for statistical
 studies of warps. It also provides a source guide for detailed
 particular observations. We confirm the large frequency of warped
 spirals: nearly all galaxies are warped. The frequency and warp angle
 do not present important differences for the different types of
 spirals. However, no lenticular warped galaxy has been found 
 within the specified limits. This finding constitutes an important
 restriction for theoretical models. 
}

\maketitle

\section{Introduction}
 
As peripheral features, disc warps are better observed in 21 cm. This
has been the preferred observational technique since their discovery by
Sancisi (\cite{sancisi})  and the study by Bosma (\cite{bosma}) until
more recent samples such as those analyzed by Garc\'{\i}a-Ru\'{\i}z
(\cite{garciaruiz01a}) and Garc\'{\i}a-Ru\'{\i}z et al.
(\cite{garciaruiz01b}). Optical observations provide an important
complementary tool. Even if the relation between optical and radio warps
remains unclear (Garc\'{\i}a-Ru\'{\i}z \cite{garciaruiz01a}), the
great advantage of optical observations lies in the much larger
samples available. Catalogs of warped galaxies have been useful to
establish observational restrictions to explain
warps. S\'anchez-Saavedra et al.
(\cite{sanchez-saavedra}) first produced a catalog of 42 optical warps
in the northern hemisphere out of a sample of 86 galaxies
analyzed. The most noticeable result was that, taking into account the
probability of non-detection of warps when the line of nodes lies in
the plane of the sky, nearly all galaxies are warped, confirming the
suggestion made by Bosma (\cite{bosma}) for HI warps. Warps are
therefore a universal feature, common for nearly all spiral
galaxies. Even this large frequency is a severe restriction for
theoretical models. As Reshetnikov and Combes said : ``Differential
precession wraps any warp'', in contrast with the large frequency of
real warped galaxies. 

Reshetnikov \& Combes (\cite{reshetnikov99}) studied 540 edge-on
galaxies, from which a sub-sample of 60 warped galaxies was
presented. One of the most noticeable new results reported in their work was
that warps were more frequent in denser environments. They also found
that the warping mechanism is equally efficient in all types of
spirals. 

Detailed studies of particular warped galaxies in the optical, such as
those by Florido et al. (\cite{florido}) and by de Grijs
(\cite{grijs}), the latter including 44 galaxies, are also of great
interest, evidently , but the production of catalogs has greater
statistical power. 
	
Here, we have examined all the galaxies contained in LEDA, with logr25
(log10 of axis ratio (major/minor axis)) $>$ 0.60, $B_{t}$ (total
B-magnitude) $<$ 14.5, $\delta$(2000) $<$ 0$^o$, -2.5 $<$ t
(morphological type code) $<$ 7. The number of galaxies fulfilling
these requirements is 276, which is our basis for this work, in which
the galaxies were analyzed by means of the DSS. Only discs with a warp
angle \emph{wa\/} (measured from the galactic center to the last
observable point with respect to the galactic plane) larger than 4$^o$
are considered to be warped. 

To determine the warp angle, \emph{wa}, and other geometrical
parameters of a warp, or even the very existence of a warp, subjective
criteria were used in previous catalogs and are in turn used in the
present one. Objective procedures
(Jim\'enez-Vicente \cite{jimenez98a}, Jim\'enez-Vicente et
al. \cite{jimenez98b}) present some problems when applied
to a very large number of galaxies and were disregarded. More
specifically, in some galaxies it was necessary to apply the automatic
star removing procedure to an excessively large number of stars,
rendering the map highly distorted. Also the separation of real warps
and spiral arms was difficult to define, even for galaxies where this
separation is clear to the human eye.   

Our work presents 150 warps extracted from 276 edge-on galaxies. The
recent study by Reshetnikov \& Combes (\cite{reshetnikov98},
\cite{reshetnikov99}) extracted 60 warps from 540 edge-on
galaxies. It is evident that these authors used
stricter criteria to define when a galaxy is definitely warped. Our
sample and scope are complementary. The Reshetnikov \&
Combes sample is limited by coordinates 0$^h$ $\leq$ $\alpha$ (1950) $\leq$
14$^h$, $\delta$ (1950) $\leq$ -17.5$^o$ ; ours by $\delta$(2000)
$\leq$ 0$^o$ only. That implies that the study by Reshetnikov \& Combes
covers 17\% of the whole sky, whereas ours covers 50\%. This, however, is
not strictly true because a large part of the Southern Sky is covered
by the Zone of Avoidance due to the large extinction near the galactic
plane. Taking into account this Avoidance Zone (-15$^o$ $\leq$ b
$\leq$ 15$^o$) our sample would cover about 40\% of the whole
sky. Even if based on a smaller number of edge-on galaxies, this large
coverage renders the present study more useful for certain statistical
tasks, such as the distribution of the orientation of warps. 
 
Another important difference between the work by Reshetnikov \&
Combes and the present study is that the former only pays attention to
the discs of spiral galaxies while ours includes lenticular galaxies. This
inclusion is very important from the theoretical point of view
because, roughly speaking, lenticulars have a disc but not gas; in
other words, the distribution of stars in lenticulars is very similar
to that in spiral galaxies. This is an overgeneralisation, because even
if the amount of gas in lenticulars is less than in even late-type
spirals, lenticulars do process some gas in the outer
parts. Another important difference between lenticulars
and spirals is that the former have a dominant bulge. This fact makes
analysis less simple, as huge bulges could in principle affect the
formation of warps. In any case observations of warps in lenticulars
have yet to be made and may introduce decisive
restrictions on the mechanisms behind warps. 

A large number of theories can be found in the literature:
non-spherical dark halos misaligned with the disc (Sparke \&
Casertano \cite{sparke}; Dubinski \& Kuijken \cite{dubinski}), gas
infall into the dark matter halo (Ostriker \& Binney
\cite{ostriker}; Binney \cite{binney}) or directly into the disc
(L\'opez-Corredoira et al. \cite{lopez-corredoira}), interactions with
companions or small satellites (Weinberg \cite{weinberg}),
intergalactic magnetic fields (Battaner et al. \cite{battaner}),
etc. Kuijken \& Garc\'{\i}a-Ru\'{\i}z (\cite{kuijken}) recently
presented a concise review summarizing several
mechanisms proposed to explain warps.

The large fraction of warped galaxies seems to exclude one of the most
obvious models, based on interactions, but this hypothesis has been
reconsidered by Weinberg (\cite{weinberg}), assuming a strong tidal
amplification by the gravitational wake of a satellite, although this fact was
not confirmed by Garc\'{\i}a-Ru\'{\i}z (\cite{garciaruiz01a}) who also
argued that the warp of the Milky Way induced by the
Magellanic Clouds should have been predicted to have a very different
orientation from that observed. Warps are common in
merging systems (Schwarzkopf \& Dettmar \cite{schwarzkopf}) but it
remains unclear whether mergers or interactions can explain all, or at
least most, warps. 

Warps are more frequent in denser environments
(Reshetnikov \& Combes \cite{reshetnikov99}). Early $z \approx 1$
warps were considered by Reshetnikov et
al. (\cite{reshetnikov2002}). Early warps were larger, which favors the
interaction model.  Other models cannot be completely excluded from
the observation of early $z \approx 1$ warps. Magnetic fields were
also much stronger and the rate of infall of matter onto a
galaxy was higher. 

Garc\'{\i}a-Ru\'{\i}z (\cite{garciaruiz01a}) observed that, even if
galaxies with extended discs may be warped, extended discs are less
frequent in denser environments. 

The observational study of warps in lenticular galaxies is
crucial. If warps are absent or are less common in lenticular galaxies
which are gas-poor, those models based on gravitation alone would have
difficulty in explaining this fact. Models in which a permanent torque
acts on the gas of the galaxy would have the preference. For
instance, models like that by Kahn \& Woltjer (\cite{kant}) or its
more recent version by L\'opez-Corredoira et
al. (\cite{lopez-corredoira}) would be favored. Note
that neither model requires the assumption that galaxies have a
large dark matter halo. The magnetic model, in which
intergalactic magnetic fields maintain the warp structure, would
acquire additional support from this fact. 

It therefore seems that the compilation of large samples of warped
galaxies, even if they contain just a few parameters about their position,
the parent galaxy and the magnitude and shape of the warping,
contributies as much as the detailed examination of the HI maps of a
few galaxies.

\section{The new catalog}

The catalog presented here is shown in Table 1. Col. 1 gives
the pgc number, Col. 2 the galaxy-name or the alternate name, Cols.
3 and 4 the source position for the epoch J2000, the right ascension
(al2000, hours decimal value) and declination (de2000, degrees
decimal value), Col. 5 the log10 of apparent diameter (d25 in 0.1'),
Col. 6 the log10 of the axis ratio (major axis/minor axis) and Col. 7 the
morphological type code (t), (directly adopted from LEDA). Cols. 8,
9, 10 and 11 are the result of our analysis, with the following meaning:  

Col. 8, labelled \emph{N/S}, specifies the apparent warp rotation,
\emph{S} clockwise, \emph{N} counterclockwise. The \emph{N} and
\emph{S} shapes are really two sides of the same galaxy. This
difference is therefore completely unimportant from the physical point
of view. However, we have kept the entries information because it is needed
when studying the distribution of warps in space, for instance, when
considering the orientation of warps in a cluster of galaxies (see for
instance, Battaner et al. 1991)  \emph{L} means that
only one of the two sides of the galaxy is warped. \emph{U} means that
the two warps are not antisymmetric, i.e. that the apparent warp
rotation on the two sides of the edge-on galaxies has opposite
directions. In this column, ``-''means unwarped and ``?'' unclear. 

Col. 9, labelled \emph{(wa)E-W}, gives the warp angle, defined as the
angle between the outermost detected point and the mean position of
the plane of symmetry, as defined by the internal unwarped
region. \emph{E}  indicates the side of the galaxy closer to the
east. \emph{W}, the side of the galaxy closer to the west. In this
column, c indicates the presence of noticeable corrugations, and b means
that the observed apparent warps could actually be arms. These are not
included as true warps.  

The angle $\beta$ in Col. 10 is the angle between the
outermost detected point and the point where the warp starts (see
Fig. 1).

\begin{figure}
   \centering
   \includegraphics[width=8cm]{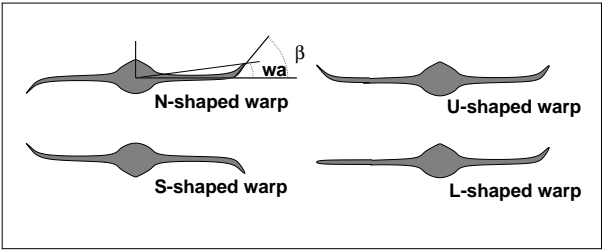}
   \caption{Definitions of angles and types of warps.}
\end{figure}

Finally, Col. 11 gives $\alpha_{s}$, the degree of asymmetry, defined as
 
\begin{equation}
 \alpha_{s} = \frac{|wa(E)-wa(W)|}{wa(E)+wa(W)}\
\end{equation}

The catalog for lenticular galaxies is presented independently in
Table \ref{tbl:lenti}. In this case, we have considered 26 galaxies
with the same limits as those used for spiral galaxies. In addition,
galaxies with logr25 $>$ 0.57 were considered. This enlarged the
sample by another 12 lenticulars. This sample is complete with the
above-mentioned limits.

\section{Basic results}

Fig. 2 gives the distribution of types of warped spiral galaxies,
together with the distribution of types for the complete (warped +
unwarped) sample. Neither differs significantly from the general
distribution of all spirals. The two distributions are so
similar, differing only in the size of the sample, that it can be
clearly concluded that for spiral galaxies the frequency of warps is
completely independent of the type. 

The degree of warping is independent of the type, both as defined by
the warp angle  \emph{(wa)} (see Fig. 3) and the angle $\beta$
(see Fig. 4). This important property was pointed out by
Reshetnikov \& Combes (\cite{reshetnikov99}). 

\begin{figure}
   \centering
   \includegraphics[width=8cm]{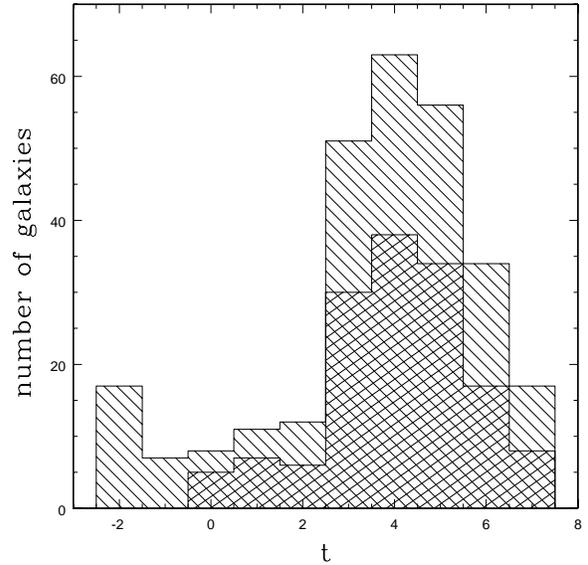}
   \caption{Distribution of types of warped spiral and lenticular galaxies. Lined shadow area represents all galaxies (warped and unwarped). Squared shadow area represents warped galaxies}
\end{figure}
  
\begin{figure}
   \centering
   \includegraphics[width=8cm]{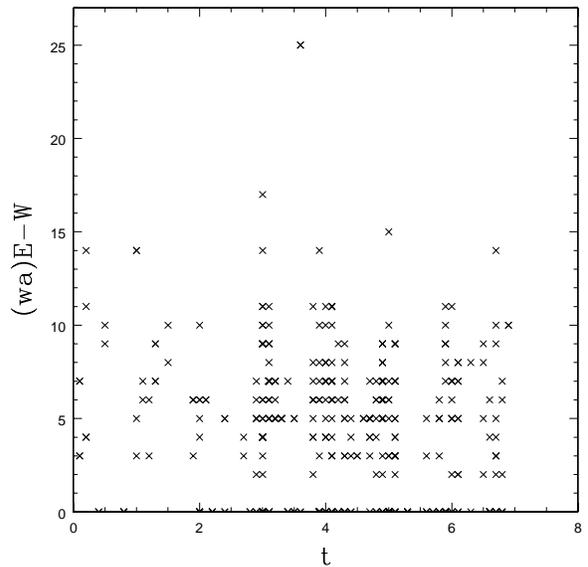}
   \caption{Warp angle versus  type.}
\end{figure} 
 
\begin{figure}
   \centering
   \includegraphics[width=8cm]{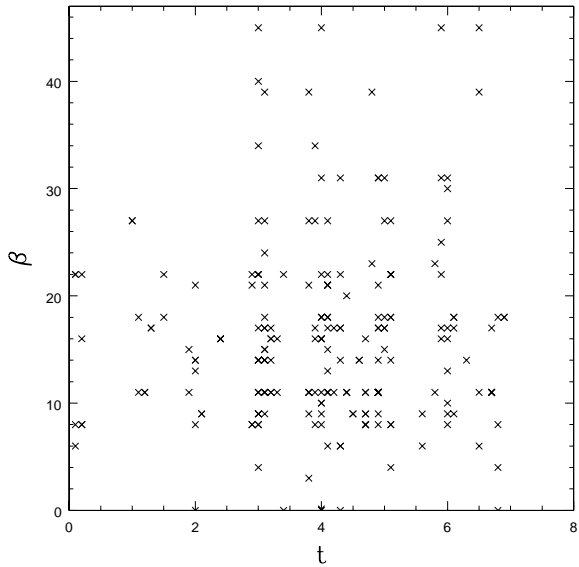}
   \caption{Angle $\beta$ versus type.}
\end{figure}  

In the case of lenticular galaxies, the result is noticeably different,
however. None of the 38 lenticular galaxies in our sample is
warped. Warping distorts discs with a
similar frequency and amount for t $>$ 0, but the transition at t=0 is
very sharp for this feature. This fact could be
interpreted in two ways.

Firstly, the lenticular dominant bulges could hamper the formation of
warps. There is no obvious theoretical argument favoring this
interpretation. On the other hand, the fact that all types of spirals
have the same warp frequency and as the size of the bulge is a
decreasing function of type, the bulge mass seems to have a small
influence on the magnitude of the warp. Therefore, it might be
suggested that for a galaxy to
be warped it must have large amounts of gas.

Fig. 5 shows the relation between \emph{wa} and $\beta$, which
gives some geometrical characteristics of the warp. For
moderate warps, there is a clear correlation between \emph{wa} and
$\beta$, as expected. However, for very large values of $\beta$,
\emph{wa} remains constant. Actually, a threshold value for \emph{wa}
seems to exist at around 14$^0$. The farther away the warp starts, the
steeper it rises. Theoretical work should pay attention to this fact.

\begin{figure}
   \centering
   \includegraphics[width=8cm]{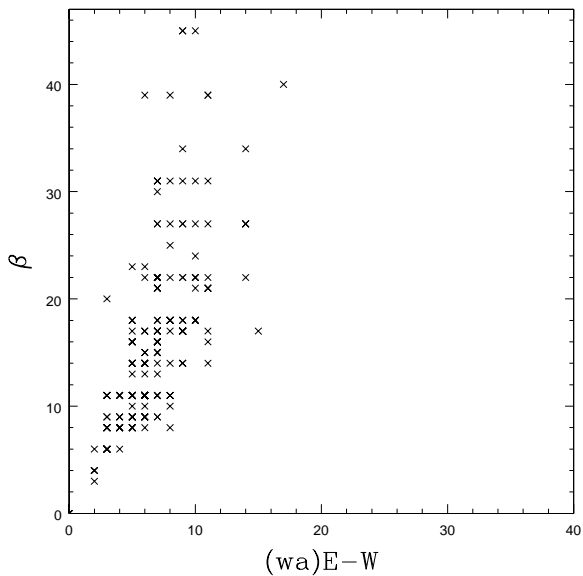}
   \caption{Relation between \emph{wa} and $\beta$.}
\end{figure}

Asymmetry seems to be unrelated to the morphological type. Fig.
6 shows that no large asymmetric warps are found in early types,
but as the frequency of these early types is much lower, this
result cannot be considered significant. 

\begin{figure}
   \centering
   \includegraphics[width=8cm]{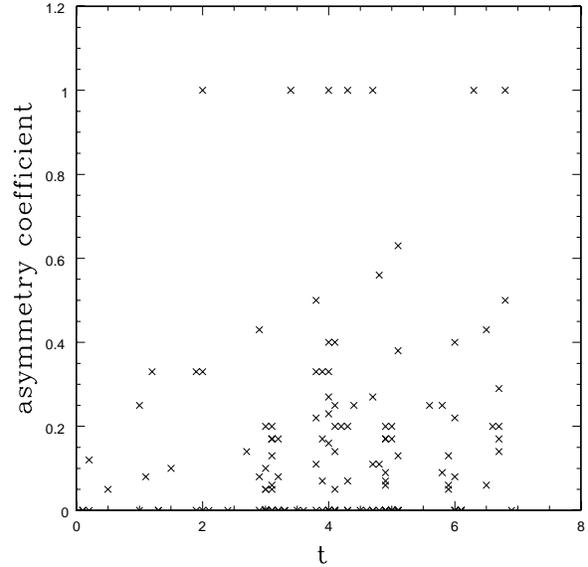}
   \caption{Asymmetry coefficient versus type.}
\end{figure}  

The frequency of warps is summarized in the following table: \\

\begin{tabular}{|cccl|}
\hline
\multicolumn{4}{|l|} {\emph{Total frequency of warped spiral galaxies}}\\
\hline
\hline
\multicolumn{4}{|l|} {Total frequency of warps, 60$\%$}\\
\multicolumn{4}{|l|} {Within warped galaxies, the frequencies are:}\\ 

\emph{N} & \emph{S}  & \emph{L} & \emph{U} \\
50$\%$ & 38$\%$ & 5$\%$ & 7$\%$ \\
\hline
\end{tabular}
\\

The \emph{N} and \emph{S} frequencies are similar, as expected, as these
characteristics depend on the observer rather than on the
galaxy. The difference, 50$\%$ and 38$\%$, is not significant (it is
within the statistical errors). The
frequency of \emph{U} warps is a parameter of theoretical
importance, however, the different scenarios predicting different values. As
stated by Castro-Rodr\'{\i}guez et al. (\cite{nieves}) an \emph{L}
warp, or even any asymmetric warp can be interpreted as a
(\emph{N+U}) or as a (\emph{S+U}) warp. 

\section{Conclusions}

   \begin{enumerate}
      \item This catalog contains a large sample of warped galaxies
      (150), covering the complete southern hemisphere. We present the
      whole sample of 250 spirals and 26 lenticulars, with limits
      logr25 $>$ 0.60, $B_{t} <$ 14.5, $\delta$ $<$ 0$^o$, -2.5 $<$ t
      $<$ 7. It is especially suitable for statistical analysis and,
      indeed, has already been used to study the relation between
      intrinsic parameters and warps by Castro-Rodr\'{\i}guez et
      al. (\cite{nieves}). The catalog may obviously be used to choose which
      galaxies are to be observed in detail.
      \item There is the unavoidable problem of the existence of other
      contaminant effects that could present the appearance of
      warps. Reshetnikov \& Combes (\cite{reshetnikov98})
      estimated that about 20$\%$ of the features assumed to be warps
      could actually be spiral arms. This value could be applied
      here. This effect introduces small errors into our calculated
      frequencies, which should be taken into account in statistical
      studies.   
      \item We confirm that warps appear to be a universal feature in
      spiral galaxies. The frequency of warps is very high (60\%), and
      because of the difficulty (or impossibility) of detecting warps with
      the line of nodes in the plane of sky, it is concluded that most
      (if not all) galaxies are warped. Bosma (\cite{bosma}),
      S\'anchez-Saavedra et al. (\cite{sanchez-saavedra}) and
      Reshetnikov \& Combes (\cite{reshetnikov99}) previously reached
      this conclusion.   
      \item We also confirm, by means of a larger amount of data, the
      finding by Reshetnikov \& Combes (\cite{reshetnikov99}) that
      warps are equally present in all types of spirals. The
      distribution of warps for the different types of spirals
      coincides with the distribution of galaxies with type. The
      maximum observed warp angle either from the center (\emph{wa})
      or from the starting radius of the warps ($\beta$) has no
      relation with the type of spiral. We have observed that very large
      values of $\beta$ do not correspond to large warp angles of
      \emph{wa}. Shorter warps are steeper. 
     \item We found no warped lenticular galaxies at all among the
      26 galaxies within our limits, with logr25 $>$ 0.60. We
      enlarged the sample to reach logr25 $>$ 0.57 and again,
      none of the 38 lenticulars was warped. There is a sudden
      transition at t=0. All spirals are warped; no lenticular is
      warped. The main difference between spirals and lenticulars is
      probably that the former are gas rich and the latter gas
      poor. Gas seems to be a necessary ingredient in the warp
      mechanism. Models based on gravitation alone would have
      serious difficulties in explaining this. 
   \end{enumerate}

\begin{acknowledgements}
      We acknowledge the use of the LEDA database
(http://leda.univ-lyon1.fr) and the Digitized Sky Survey (DSS) of
NASA's SkyView facility (http://skyview.gsfc.nasa.gov) located at NASA
Goddard Space Flight Center.

\end{acknowledgements}

\begin{table*}[ht]
\caption{Sample of Spiral Galaxies: In the 8th column
 ``-''means unwarped and ``?'' unclear; in the 9th column ``b'' means arms,
 ``c'' means corrugated and ``0'' means unwarped.}
\begin{tabular}{ccccccccccc}
\hline
pgc & galaxy-name & al2000 & de2000 & logd25 &	logr25 & t &
\emph{N/S} & \emph{(wa)E-W} ($^o$) & $\beta$ ($^o$) & $\alpha_{s}$ \\
\hline
43 & ESO293-27 & 0.00817 & -40.48400 & 1.32 &	0.60 & 3.9 &	LS & ? & - & - \\
474 & MCG-7-1-9 & 0.10585 & -41.48350 & 1.53 &	0.62 & 6.0 &	N & 7-7 & 30-31 & 0 \\
627 & NGC7 & 0.13899 & -29.91700 & 1.39 &	0.67 & 4.9 &	U & 7-7 & 31-31 & 0 \\
725 & ESO241-21 & 0.17223 & -46.41940 & 1.34 &	0.62 & 3.1 &	N & 11-10 & 39-24 & 0.05 \\
& \\
1335 & ESO78-22 & 0.34918 & -63.85740 & 1.34 &	0.79 & 4.3 &	S & 6-9 & 17-31 & 0.2 \\
1851 & NGC134 & 0.50616 & -33.24550 & 1.92 &    0.64 & 4.0 &	N & 11-8 & 22-31 & 0.16 \\
1942 & ESO473-25 & 0.53050 & -26.72010 & 1.41 &	0.84 & 4.9 &	N & 5-7 & - & 0.17 \\
1952 & ESO79-3 & 0.53390 & -64.25240 & 1.45 &	0.72 & 3.1 &	N & 9-7 & 27-21 & 0.13 \\
&\\
2228 & NGC172 & 0.62067 & -22.58500 & 1.31 &	0.72 & 4.0 &	S & 7-7 & - & 0 \\
2482 & NGC217 & 0.69267 & -10.02120 & 1.45 &   	0.65 & 0.5 &	N & 9-10 & - & 0.05 \\
2789 & NGC253 & 0.79252 & -25.28840 & 2.43 &   	0.66 & 5.1 &	N & 7-9 & 22-27 & 0.13 \\
2800 & MCG-2-3-15 & 0.79624 & -9.83460 & 1.24 &	0.64 & 3.1 &	U & 7-7 & - & 0 \\
&\\
2805 & MCG-2-3-16 & 0.79641 & -9.89970 & 1.48 &	0.82 & 6.7 &	S & 14-10 & - & 0.17 \\
2820 & NGC259 & 0.80091 & -2.77630 & 1.46 &    	0.65 & 4.0 &	S & 7-9 & 45-45 & 0.13 \\
3743 & NGC360 & 1.04763 & -65.60990 & 1.57 &   	0.89 & 4.3 &	N & 8-7 & 22-17 & 0.07 \\
4440 & IC1657 & 1.23536 & -32.65060 & 1.41 &	0.65 & 3.8 &   	S & 7-11 & 21-39 & 0.22 \\
&\\
4912 & ESO476-5 & 1.35492 & -22.80050 & 1.29 &	0.61 & 3.9 &	N & 10-14 & 27-34 & 0.17 \\
5128 & NGC527 & 1.39947 & -35.11520 & 1.23 &   	0.62 & 0.1 &	N & 7-7 & 22-22 & 0 \\
5688 & NGC585 & 1.52839 & -0.93300 & 1.34 &	0.66 & 1.0 &	N & 5-3 & - & 0.25 \\
6044 & ESO297-16 & 1.63329 & -40.06770 & 1.20 &	0.61 & 5.9 &	N & 9-7 & 17-16 & 0.13 \\
&\\
6161 & ESO413-16 & 1.66594 & -28.69710 & 1.29 &	0.67 & 3.9 &	S & 8-7 & 17-16 & 0.07 \\
6242 & ESO3-4 & 1.69266 & -83.21210 & 1.28 &	0.70 & 4.1 &	S & 7-7 & 22-27 & 0 \\
6966 & MCG-1-5-47 & 1.88028 & -3.44720 & 1.48 &	0.87 & 4.8 &	S & 2-7 & - & 0.56 \\
7298 & ESO245-10 & 1.94570 & -43.97350 & 1.34 &	0.61 & 3.2 &	? & ? & - & - \\
&\\
7306 & IC176 & 1.94814 & -2.01910 & 1.29 &	0.68 & 5.1 &	N & 7-7 & 22-22 & 0 \\
7427 & ESO297-37 & 1.97033 & -39.54460 & 1.27 &	0.67 & 4.1 &	N & 3-7 & - & 0.4 \\
8326 & ESO30-9 & 2.17801 & -75.03880 & 1.40 &	0.61 & 4.8 &	S & 4-5 & - & 0.11 \\
8581 & MCG-1-6-77 & 2.24055 & -7.36840 & 1.36 &	0.78 & 2.7 &	? & ? & - & - \\
&\\
8673 & IC217 & 2.26957 & -11.92660 & 1.34 &	0.67 & 5.8 &	U & 5-6 & - & 0.09 \\
9582 & NGC964 & 2.51835 & -36.03420 & 1.33 &	0.60 & 1.9 &	S & 3-6 & - & 0.33 \\
10645 & ESO546-25 & 2.81346 & -19.97110 & 1.16 & 0.73 & 3.6 &	- & 0-0 & - & - \\
10965 & NGC1145 & 2.90931 & -18.63550 & 1.47 &	0.74 & 5.1 &	N & ?-4 & - & - \\
&\\
11198 & MCG-2-8-33 & 2.96379 & -10.16720 & 1.45 & 0.63 & 0.2 &	N & 14-11 & 22-16 & 0.12 \\
11595 & ESO248-2 & 3.08523 & -45.96350 & 1.47 &	0.69 & 6.9 &	S & 10-10 & 18-18 & 0 \\
11659 & NGC1244 & 3.10866 & -66.77640 & 1.26 &	0.63 & 2.1 &	S & 6-6 & 9-9 & 0 \\
11851 & IC1898 & 3.17224 & -22.40440 & 1.53 &	0.77 & 5.8 &	N & 5-3 & 23-11 & 0.25 \\ 
&\\
11931 & NGC1247 & 3.20399 & -10.48070 & 1.54 &	0.82 & 3.8 &	N & 4-5 & - & 0.11 \\
12521 & NGC1301 & 3.34317 & -18.71590 & 1.34 &	0.66 & 4.2 &	N & 9-6 & 17-11 & 0.2 \\
13171 & IC1952 & 3.55728 & -23.71280 & 1.40 &	0.66 & 4.1 &	N & 8-6 & 18-13 & 0.14 \\
13458 & NGC1406 & 3.65626 & -31.32200 & 1.59 &	0.70 & 4.4 &	S & 3-5 & 20-11 & 0.25 \\
&\\
13569 & NGC1422 & 3.69191 & -21.68240 & 1.36 &	0.63 & 2.4 &	- & 0-0 & - & - \\
13620 & NGC1421 & 3.70818 & -13.48830 & 1.54 &	0.60 & 4.1 &	? & ? & - & - \\
13646 & MCG-2-10-9 & 3.71559 & -12.91570 & 1.50 & 0.90 & 5.0 &	N & 7-5 & 27-18 & 0.17 \\
13727 & NGC1448 & 3.74221 & -44.64400 & 1.88 &	0.65 & 5.9 &	S & 9-8 & 45-25 & 0.06 \\
&\\
13809 & ESO358-63 & 3.77189 & -34.94240 & 1.69 & 0.64 & 4.7 &	LS & 3-0 & - & 1 \\
13912 & IC2000 & 3.81879 & -48.85810 & 1.63 &	0.74 & 6.1 &	N & 5-5 & - & 0 \\
13926 & ESO482-46 & 3.82835 & -26.99350 & 1.55 & 0.82 & 5.1 &	? & c & - & - \\
14071 & NGC1484 & 3.90487 & -36.97110 & 1.40 &	0.69 & 3.5 &	S & 5-5 & - & 0 \\
\hline
\end{tabular}
%\label{tbl:spiral}
\end{table*}
\newpage
\begin{table*}[ht]
\caption{ (Table 1, continuation)}
\begin{tabular}{ccccccccccc}
\hline
pgc & galaxy-name & al2000 & de2000 & logd25 &	logr25 & t &
\emph{N/S} & \emph{(wa)E-W} ($^o$) & $\beta$ ($^o$) & $\alpha_{s}$\\
\hline
14190 & NGC1495 & 3.97255 & -44.46650 & 1.46 &	0.73 & 5.1 &	S & 4-9 & - & 0.38 \\
14255 & NGC1511A & 4.00542 & -67.80730 & 1.28 &	0.63 & 1.3 &	S & 7-7 & - & 0 \\
14259 & ESO483-6 & 4.00724 & -25.18150 & 1.43 &	0.79 & 3.2 &	S & 5-7 & - & 0.17 \\
14337 & ESO117-19 & 4.04233 & -62.31570 & 1.31 & 0.76 & 3.9 &	- & 0-0 & - & - \\
&\\
14397 & NGC1515 & 4.06748 & -54.10270 & 1.72 &	0.62 & 4.0 &	N & 5-8 & 10-10 & 0.23 \\
14824 & IC2058 & 4.29847 & -55.93440 & 1.49 &	0.94 & 6.6 &	S & 4-6 & - & 0.2 \\
15455 & ESO202-35 & 4.53769 & -49.67520 & 1.44 & 0.81 & 3.3 &	N & 5-5 & 16-11 & 0 \\
15635 & NGC1622 & 4.61015 & -3.18920 & 1.52 &	0.73 & 2.0 &	S & 5-10 & 13-21 & 0.33 \\
&\\
15654 & NGC1625 & 4.61841 & -3.30340 & 1.37 &	0.64 & 3.1 &	S & 6-6 & 11-? & 0 \\
15674 & NGC1628 & 4.62671 & -4.71480 & 1.27 &  	0.64 & 3.1 &	N & 8-9 & 11-17 & 0.06 \\
15749 & ESO157-49 & 4.66038 & -53.01200 & 1.28 & 0.66 & 4.1 &	S & 11-11 & 21-21 & 0 \\
15758 & IC2103 & 4.66323 & -76.83680 & 1.28 &	0.76 & 4.9 &	? & b & - & - \\
&\\
16144 & IC2098 & 4.84561 & -5.41870 & 1.35 &	0.87 & 5.9 &	- & 0-0 & - & - \\
16168 & MCG-1-13-22 & 4.85739 & -3.12210 & 1.22 & 0.60 & 3.0 &	S & 4-4 & - & 0 \\
16187 & IC2101 & 4.86165 & -6.22970 & 1.23 &	0.66 & 5.0 &	? & ? & - & - \\
16199 & ESO361-15 & 4.86601 & -33.17860 & 1.44 & 0.68 & 6.1 &	S & 7-7 & 9-17 & 0 \\
&\\
16239 & NGC1686 & 4.88184 & -15.34620 & 1.25 &	0.71 & 4.0 &	N & 4-7 & --16 & 0.27 \\
16636 & MCG-1-13-50 & 5.05475 & -2.93560 & 1.39 & 0.82 & 3.0 &	- & 0-0 & - & - \\
16849 & NGC1827 & 5.16768 & -36.95890 & 1.48 &	0.76 & 5.9 &	? & - & - & - \\
16893 & MCG-1-14-3 & 5.19493 & -3.09160 & 1.16 & 0.67 & 3.1 &	N & 7-5 & - & 0.17 \\
&\\
17027 & ESO362-11 & 5.27748 & -37.10210 & 1.68 & 0.75 & 4.1 &	N & 6-4 & 11-6 & 0.2 \\
17056 & IC407 & 5.29517 & -15.52370 & 1.24 &   	0.70 & 4.9 &	S & 6-6 & - & 0 \\
17174 & NGC1886 & 5.36352 & -23.81260 & 1.51 &	0.77 & 3.9 &	- & 0-0 & - & - \\
17248 & MCG-2-14-16 & 5.41515 & -12.68870 & 1.28 & 0.77 & 2.0 & - & 0-0 & - & - \\
&\\
17433 & NGC1963 & 5.55355 & -36.39980 & 1.46 &	0.78 & 5.8 &	? & ? & - & - \\
17969 & ESO555-2 & 5.84077 & -19.72620 & 1.34 &	0.74 & 3.9 &	S & 6-6 & - & 0 \\
17993 & ESO160-2 & 5.85418 & -53.57480 & 1.26 &	0.63 & 3.0 &	? & ? & - & - \\
18394 & ESO5-4 & 6.09456 & -86.63220 & 1.58 &	0.68 & 2.9 &	- & 0-0 & - & - \\
&\\
18437 & ESO121-6 & 6.12505 & -61.80710 & 1.60 &	0.75 & 5.1 &	S & 3-3 & - & 0 \\
18765 & ESO489-29 & 6.28479 & -27.38620 & 1.52 & 0.72 & 3.9 &	- & 0-0 & - & - \\
18833 & NGC2221 & 6.33750 & -57.57740 & 1.33 &	0.66 & 1.1 &	U & 7-6 & 18-11 & 0.08 \\
19996 & ESO491-15 & 7.01183 & -27.36820 & 1.35 & 0.68 & 5.0 & 	N & 3-3 & - & 0 \\
&\\
20903 & ESO428-28 & 7.39406 & -30.05070 & 1.39 & 0.84 & 5.3 &	U & - & - & - \\
21338 & ESO257-19 & 7.58538 & -46.92470 & 1.40 & 0.74 & 6.2 &	? & b & - & - \\
21815 & ESO311-12 & 7.79281 & -41.45170 & 1.57 & 0.74 & 0.1 &	S & 3-3 & 6-8 & 0 \\
21822 & ESO560-13 & 7.79774 & -18.74840 & 1.48 & 0.77 & 4.0 &	N & 10-4 & 18-8 & 0.4 \\
&\\
22174 & ESO35-18 & 7.91809 & -76.41310 & 1.53 &	0.65 & 4.9 &	- & 0-0 & - & - \\
22272 & ESO494-7 & 7.94852 & -24.90810 & 1.45 &	0.63 & 4.3 &	- & 0-0 & - & - \\
22338 & ESO209-9 & 7.97080 & -49.85160 & 1.80 &	0.83 & 6.0 &	N & 5-5 & 9-17 & 0 \\
22910 & ESO89-12 & 8.16682 & -64.93620 & 1.41 &	0.86 & 4.1 &	N & 7-7 & 15-17 & 0 \\
&\\
23558 & ESO495-12 & 8.39769 & -25.83820 & 1.27 & 0.67 & 3.0 &	S & 11-11 & 14-17 & 0 \\
23672 & IC2375 & 8.43881 & -13.30300 & 1.28 &	0.70 & 3.0 &	S & 17-14 & 40-27 & 0.1 \\
23992 & ESO562-14 & 8.55493 & -17.95650 & 1.15 & 0.64 & 3.0 &	N & 9-10 & 34-45 & 0.05 \\
23997 & NGC2613 & 8.55626 & -22.97330 & 1.85 &	0.62 & 3.2 &	? & c & - & - \\
&\\
24225 & ESO563-3 & 8.62196 & -20.93980 & 1.21 &	0.65 & 0.4 &	- & 0-0 & - & - \\
24479 & ESO563-14 & 8.71596 & -20.05070 & 1.38 & 0.66 & 6.5 &	? & c & - & - \\
24685 & ESO563-21 & 8.78801 & -20.03590 & 1.48 & 0.81 & 4.1 &	S & 3-5 & - & 0.25 \\
25400 & ESO60-24 & 9.04451 & -68.22650 & 1.46 &	0.78 & 2.9 &	S & 2-5 & - & 0.43 \\
\hline
\end{tabular}
%\label{tbl:spiral}
\end{table*}
\newpage
\begin{table*}[ht]
\caption{ (Table 1, continuation)}
\begin{tabular}{ccccccccccc}
\hline
pgc & galaxy-name & al2000 & de2000 & logd25 &	logr25 & t &
\emph{N/S} & \emph{(wa)E-W} ($^o$) & $\beta$ ($^o$) & $\alpha_{s}$\\
\hline
25886 & MCG-1-24-1 & 9.18087 & -8.88820 & 1.63 & 0.63 & 3.3 &	S & 5-5 & - & 0 \\
25926 & ESO564-27 & 9.19844 & -20.11760 & 1.65 & 0.94 & 6.1 &	S & 2-2 & - & 0 \\
26561 & IC2469 & 9.38368 & -32.45070 & 1.75 &	0.68 & 1.9 &	N & 6-6 & 11-15 & 0 \\
26632 & ESO433-19 & 9.40089 & -28.17720 & 1.15 & 0.76 & 0.8 &	- & 0-0 & - & - \\
&\\
27135 & ESO373-8 & 9.55577 & -33.03250 & 1.76 &	0.81 & 6.0 &	- & 0-0 & - & - \\
27468 & ESO373-13 & 9.63908 & -33.86110 & 1.13 & 0.81 & 0.2 &	U & 4-4 & 8-8 & 0 \\
27735 & MCG-1-25-22 & 9.70339 & -4.71400 & 1.28 & 0.66 & 4.0 &	N & 7-7 & 16-16 & 0 \\
27982 & NGC2992 & 9.76169 & -14.32750 & 1.56 &	0.62 & 1.0 &	? & ? & - & - \\
&\\
28117 & ESO499-5 & 9.78709 & -24.84040 & 1.41 &	0.64 & 5.0 &	S & 10-15 & 31-17 & 0.2 \\
28246 & IC2511 & 9.82372 & -32.84200 & 1.50 &	0.65 & 1.5 &	N & 8-10 & 18-22 & 0.1 \\
28283 & IC2513 & 9.83403 & -32.88580 & 1.49 &	0.60 & 2.2 & 	- & 0-0 & - & - \\
28308 & MCG-2-25-20 & 9.83718 & -12.05720 & 1.25 & 0.78 & 6.8 &	LS & 7-0 & 18-- & 1 \\
&\\
28778 & ESO435-14 & 9.96341 & -28.50670 & 1.40 & 0.86 & 4.9 &	N & 5-5 & 11-14 & 0 \\
28840 & ESO435-19 & 9.98502 & -30.24980 & 1.53 & 0.86 & 4.7 &	U & 4-5 & 8-8 & 0.11 \\
28909 & IC2531 & 9.99880 & -29.61540 & 1.83 &	0.96 & 5.0 &	- & 0-0 & - & - \\
29096 & ESO316-18 & 10.04541 & -42.09030 & 1.40 & 0.85 & 4.9 &	S & 5-6 & - & 0.09 \\
&\\
29691 & NGC3157 & 10.19515 & -31.64220 & 1.35 &	0.62 & 5.0 &	- & 0-0 & - & - \\
29716 & ESO263-15 & 10.20551 & -47.29430 & 1.49 & 0.91 & 5.8 &	- & 0-0 & - & - \\
29743 & ESO436-1 & 10.21327 & -27.83950 & 1.50 & 0.80 & 4.3 &	N & 5-5 & - & 0 \\
29841 & ESO567-26 & 10.23436 & -21.97680 & 1.32 & 0.72 & 4.1 & 	- & 0-0 & - & - \\
&\\
30716 & ESO375-26 & 10.45064 & -36.22700 & 1.30 & 0.74 & 4.1 &	- & 0-0 & - & - \\
30887 & NGC3263 & 10.48691 & -44.12300 & 1.78 &	0.64 & 5.9 &	? & ? & - & - \\
31154 & ESO436-34 & 10.54559 & -28.61290 & 1.35 & 0.65 & 3.0 &	- & 0-0 & - & - \\
31426 & IC624 & 10.60427 & -8.33410 & 1.41 &	0.61 & 1.2 &	S & 6-3 & 11-11 & 0.33 \\
&\\
31626 & ESO437-22 & 10.63830 & -28.88600 & 1.22 & 0.63 & 4.0 &	? & c & - & - \\
31677 & ESO437-30 & 10.65422 & -30.29910 & 1.50 & 0.73 & 4.0 &	LN & 6-0 & 9-- & 1 \\
31723 & NGC3333 & 10.66387 & -36.03610 & 1.32 &	0.63 & 4.3 &	LN & 6-0 & 14-- & 1 \\
31919 & ESO501-80 & 10.71052 & -23.93550 & 1.36 & 0.67 & 5.0 &	- & 0-0 & - & - \\
&\\
31995 & ESO318-4 & 10.73068 & -38.26280 & 1.45 & 0.63 & 5.1 &	N & 5-5 & 18-18 & 0 \\
32271 & NGC3390 & 10.80110 & -31.53260 & 1.54 &	0.73 & 2.7 &	N & 3-4 & - & 0.14 \\
32328 & ESO264-43 & 10.81209 & -45.42020 & 1.30 & 0.64 & 3.1 &	- & 0-0 & - & - \\
32550 & ESO569-14 & 10.85682 & -19.88890 & 1.54 & 0.72 & 6.3 &	LN & 8-0 & 14-- & 1 \\
&\\
35539 & NGC3717 & 11.52557 & -30.30770 & 1.78 &	0.64 & 3.1 &	? & ? & - & - \\
35861 & NGC3749 & 11.59794 & -37.99470 & 1.53 &	0.60 & 1.0 &	U & 14-14 & 27-27 & 0 \\
36315 & ESO571-16 & 11.70260 & -18.16960 & 1.21 & 0.66 & 3.9 &	N & 6-3 & 11-8 & 0.33 \\
37178 & NGC3936 & 11.87235 & -26.90650 & 1.60 &	0.73 & 4.4 &	- & 0-0 & - & - \\
&\\
37243 & ESO379-6 & 11.88444 & -36.63820 & 1.42 & 0.90 & 4.9 &	N & 3-2 & - & 0.2 \\
37271 & ESO440-27 & 11.88987 & -28.55320 & 1.64 & 0.83 & 6.7 &	S & 4-3 & 11-11 & 0.14 \\
37304 & IC2974 & 11.89692 & -5.16780 & 1.36 &	0.69 & 4.7 &	S & 5-5 & 9-9 & 0 \\
37334 & ESO320-31 & 11.90167 & -39.86450 & 1.42 & 0.88 & 5.1 &	- & 0-0 & - & - \\
&\\
38426 & MCG-2-31-17 & 12.11402 & -11.09900 & 1.31 & 0.77 & 6.0 & S & 7-6 & 16-10 & 0.08 \\
38464 & IC3005 & 12.12049 & -30.02330 & 1.38 &	0.76 & 5.6 &	N & 3-5 & 6-9 & 0.25 \\
38841 & ESO321-10 & 12.19502 & -38.54850 & 1.31 & 0.83 & 0.8 &	- & 0-0 & - & - \\
40023 & ESO380-19 & 12.36725 & -35.79230 & 1.50 & 0.76 & 5.8 &	- & 0-0 & - & - \\
&\\
40284 & NGC4348 & 12.39832 & -3.44330 & 1.49 &	0.70 & 4.1 &	N & 10-11 & 18-21 & 0.05 \\
42684 & ESO268-33 & 12.70824 & -47.55780 & 1.32 & 0.71 & 4.9 &	- & 0-0 & - & - \\
42747 & UGC7883 & 12.71593 & -1.22940 & 1.42 &	0.61 & 6.0 &	N & 7-11 & 13-27 & 0.22 \\
43021 & ESO507-7 & 12.76168 & -26.24320 & 1.41 & 0.82 & 4.0 &	- & 0-0 & - & - \\
\hline
\end{tabular}
%\label{tbl:spiral}
\end{table*}
\newpage
\begin{table*}[ht]
\caption{ (Table 1, continuation)}
\begin{tabular}{ccccccccccc}
\hline
pgc & galaxy-name & al2000 & de2000 & logd25 &	logr25 & t &
\emph{N/S} & \emph{(wa)E-W} ($^o$) & $\beta$ ($^o$) & $\alpha_{s}$ \\
\hline
43224 & ESO507-13 & 12.80150 & -27.57800 & 1.25 & 0.62 & 4.1 &	N & 6-6 & 11-11 & 0 \\
43313 & IC3799 & 12.81658 & -14.39910 & 1.40 &	0.88 & 6.7 &	- & 0-0 & - & - \\
43330 & NGC4700 & 12.81883 & -11.41060 & 1.47 &	0.75 & 4.9 &	N & 8-9 & 11-18 & 0.06 \\
43342 & NGC4703 & 12.82187 & -9.10850 & 1.39 &	0.69 & 3.1 &	- & 0-0 & - & - \\
&\\
43679 & MCG-1-33-32 & 12.87411 & -9.75390 & 1.39 & 0.92 & 6.7 &	- & 0-0 & - & - \\
44254 & UGC8067 & 12.95337 & -1.70690 & 1.28 &	0.66 & 3.5 &	- & 0-0 & - & - \\
44271 & NGC4835A & 12.95364 & -46.37780 & 1.43 & 0.62 & 5.8 &	- & 0-0 & - & - \\
44358 & MCG-1-33-60 & 12.96300 & -9.63360 & 1.51 & 0.94 & 6.7 &	- & 0-0 & - & - \\
&\\
44409 & NGC4835 & 12.96883 & -46.26320 & 1.67 &	0.69 & 4.0 &	? & b & - & - \\
44931 & MCG-1-33-71 & 13.03043 & -8.33620 & 1.45 & 0.85 & 4.9 &	N & 8-7 & 8-9 & 0.07 \\
44966 & ESO381-51 & 13.03531 & -33.11870 & 1.19 & 0.70 & 2.8 &	- & 0-0 & - & - \\
45006 & MCG-3-33-28 & 13.04054 & -17.67920 & 1.42 & 0.92 & 4.9 & N & 5-7 & 11-21 & 0.17 \\
&\\
45098 & ESO443-42 & 13.05827 & -29.82900 & 1.46 & 0.83 & 3.1 &	N & 7-5 & 17-14 & 0.17 \\
45127 & MCG-1-33-76 & 13.06291 & -5.13370 & 1.27 & 0.61 & 4.9 &	N & 6-6 & 11-11 & 0 \\
45279 & NGC4945 & 13.09060 & -49.47090 & 2.31 & 0.67 & 6.1 &	N & 8-8 & 18-18 & 0 \\
45487 & ESO508-11 & 13.12910 & -22.85680 & 1.48 & 0.79 & 6.7 &	- & 0-0 & - & - \\
&\\
45911 & ESO576-11 & 13.21811 & -19.97810 & 1.47 & 0.79 & 5.7 &	- & 0-0 & - & - \\
45952 & NGC5022 & 13.22530 & -19.54800 & 1.38 &	0.69 & 3.4 &	- & 0-0 & - & - \\
46441 & NGC5073 & 13.32241 & -14.84440 & 1.54 &	0.79 & 5.0 &	S & 6-6 & 17-15 & 0 \\
46650 & ESO40-7 & 13.36095 & -77.53510 & 1.46 &	0.80 & 5.1 &	N & 2-9 & 4-14 & 0.63 \\
&\\
46768 & IC4231 & 13.38707 & -26.30050 & 1.25 &	0.65 & 4.3 & 	- & 0-0 & - & - \\
46928 & ESO382-58 & 13.42014 & -33.65550 & 1.41 & 0.73 & 3.9 &	? & c & - & - \\
47345 & ESO383-5 & 13.48987 & -34.27190 & 1.52 & 0.72 & 3.8 &	N & 4-8 & 11-27 & 0.33 \\
47394 & NGC5170 & 13.49692 & -17.96620 & 1.91 &	0.84 & 4.9 &	- & 0-0 & - & - \\
&\\
47948 & ESO509-74 & 13.59484 & -24.07400 & 1.40 & 0.72 & 4.7 &	N & 5-5 & 11-11 & 0 \\
48359 & ESO220-28 & 13.67018 & -51.14220 & 1.31 & 0.75 & 4.2 &	- & 0-0 & - & - \\
49106 & IRAS13471-4839 & 13.83850 & -48.90490 & 1.25 & 0.72 & 3.6 & S & 25-25 & ? & 0 \\
49129 & ESO383-91 & 13.84225 & -37.28920 & 1.41 & 0.81 & 6.7 &	N & 5-9 & 11-17 & 0.29 \\
&\\
49190 & ESO384-3 & 13.85615 & -37.62870 & 1.23 & 0.67 & 3.0 &	S & 10-9 & 22-14 & 0.05 \\
49586 & NGC5365A & 13.94436 & -44.00730 & 1.45 & 0.68 & 3.0 &	N & 5-5 & 11-8 & 0 \\
49676 & IC4351 & 13.96504 & -29.31490 & 1.76 &	0.69 & 3.2 & 	N & 6-7 & 11-14 & 0.08 \\
49750 & NGC5365B & 13.97766 & -43.96420 & 1.21 & 0.65 & 2.0 &	- & 0-0 & - & - \\
&\\
49788 & ESO325-42 & 13.98897 & -40.06910 & 1.21 & 0.63 & 3.2 &	S & 5-5 & - & 0 \\
49836 & ESO221-22 & 14.00347 & -48.26770 & 1.37 & 0.69 & 6.8 &	S & 2-6 & 4-8 & 0.5 \\
50676 & NGC5496 & 14.19388 & -1.15890 & 1.64 & 0.69 & 6.5 &	S & 8-9 & 39-45 & 0.06 \\
50798 & ESO271-22 & 14.22487 & -45.41360 & 1.41 & 0.73 & 5.9 &	- & 0-0 & - & - \\
&\\
51288 & IC4402 & 14.35379 & -46.29830 & 1.68 &	0.69 & 3.2 &	? & c & 17-11 & - \\
51613 & ESO1-6 & 14.45741 & -87.77160 & 1.48 &	0.64 & 5.9 &	- & 0-0 & - & - \\
52410 & IC4472 & 14.66980 & -44.31610 & 1.35 &	0.66 & 5.0 &	- & 0-0 & - & - \\
52411 & ESO512-12 & 14.66983 & -25.77610 & 1.45 & 0.82 & 3.2 &	N & 5-5 & 16-16 & 0 \\
&\\
52824 & ESO580-29 & 14.79267 & -19.76510 & 1.34 & 0.80 & 4.9 &	N & 6-6 & 11-11 & 0 \\
52991 & ESO580-41 & 14.84346 & -18.15090 & 1.30 & 0.70 & 4.3 &	- & 0-0 & - & - \\
53361 & ESO327-31 & 14.92637 & -38.27700 & 1.36 & 0.74 & 5.0 &	- & 0-0 & - & - \\
53471 & MCG-7-31-3 & 14.96256 & -43.13190 & 1.18 & 0.70 & 4.4 &	? & ? & - & - \\
&\\
54348 & ESO581-25 & 15.22503 & -20.67680 & 1.54 & 0.70 & 6.9 &	? & c & - & - \\
54392 & ESO274-1 & 15.23711 & -46.81250 & 2.05 & 0.76 & 6.6 &	- & 0-0 & - & - \\
54637 & ESO328-41 & 15.30663 & -38.50690 & 1.40 & 0.69 & 3.1 &	S & 5-7 & 9-11 & 0.17 \\
56077 & IC4555 & 15.80448 & -78.17830 & 1.30 &	0.64 & 5.9 &	- & 0-0 & - & - \\
\hline
\end{tabular}
%\label{tbl:spiral}
\end{table*}
\newpage
\begin{table*}[ht]
\caption{ (Table 1, continuation)}
\begin{tabular}{ccccccccccc}
\hline
pgc & galaxy-name & al2000 & de2000 & logd25 &	logr25 & t &
\emph{N/S} & \emph{(wa)E-W} ($^o$) & $\beta$ ($^o$) & $\alpha_{s}$ \\
\hline
57582 & UGC10288 & 16.24027 & -0.20780 & 1.68 &	0.94 & 5.3 &	- & 0-0 & - & - \\
57876 & IC4595 & 16.34586 & -70.14160 & 1.49 &	0.75 & 5.0 &	- & 0-0 & - & - \\
58742 & ESO137-38 & 16.68135 & -60.39340 & 1.48 & 0.60 & 4.4 &	? & ? & - & - \\
59635 & ESO138-14 & 17.11666 & -62.08300 & 1.57 & 0.80 & 6.7 &	- & 0-0 & - & - \\
&\\
60216 & ESO138-24 & 17.40184 & -59.38210 & 1.30 & 0.65 & 4.9 &	S & 9-9 & 17-17 & 0 \\
60595 & IC4656 & 17.62894 & -63.72950 & 1.39 &	0.61 & 5.0 & 	- & 0-0 & - & - \\
60772 & ESO139-21 & 17.73619 & -60.97850 & 1.32 & 0.60 & 3.0 &	- & 0-0 & - & - \\
62024 & IC4717 & 18.55492 & -57.97400 & 1.22 &	0.61 & 3.0 &	- & 0-0 & - & - \\
&\\
62529 & ESO281-33 & 18.88273 & -42.53750 & 1.24 & 0.61 & 3.0 &	? & ? & - & - \\
62706 & IC4810 & 19.04974 & -56.15860 & 1.57 & 	0.91 & 6.6 &	- & 0-0 & - & - \\
62722 & NGC6722 & 19.06100 & -64.89480 & 1.46 &	0.67 & 2.9 &	S & 5-5 & 8-8 & 0 \\
62782 & IC4819 & 19.11826 & -59.46550 & 1.47 &	0.73 & 6.0 &	S & 5-2 & 8-- & 0.4 \\
&\\
62816 & ESO231-23 & 19.14551 & -51.04600 & 1.25 & 0.67 & 3.0 &	U & 6-6 & 11-14 & 0 \\
62922 & IC4827 & 19.22261 & -60.86010 & 1.47 &	0.68 & 2.0 &	LN & 4-0 & 8-0  & 1 \\
62938 & IC4832 & 19.23426 & -56.60900 & 1.37 &	0.64 & 1.3 &	N & 9-9 & 17-17 & 0 \\
62964 & IC4837A & 19.25435 & -54.13250 & 1.62 &	0.73 & 3.1 &	N & 7-7 & 15-15 & 0 \\
&\\
63161 & ESO184-63 & 19.39455 & -55.06580 & 1.36 & 0.72 & 2.9 &	N & 7-6 & 21-22 & 0.08 \\ 
63297 & ESO184-74 & 19.50854 & -57.28420 & 1.25 & 0.70 & 2.9 &	? & b & - & - \\
63395 & IC4872 & 19.59511 & -57.51840 & 1.51 &	0.76 & 6.7 &	S & 2-3 & --11 & 0.2 \\
63509 & ESO142-30 & 19.67763 & -60.04800 & 1.26 & 0.67 & 4.9 &	? & b & - & - \\
&\\
63577 & IC4885 & 19.73113 & -60.65150 & 1.29 &	0.64 & 4.9 &	S & 7-7 & 11-11 & 0 \\
64180 & ESO105-26 & 20.15782 & -66.21610 & 1.18 & 0.64 & 4.2 &	? & ? & - & - \\
64240 & NGC6875A & 20.19888 & -46.14380 & 1.47 & 0.70 & 4.2 &	? & b & - & - \\
64597 & IC4992 & 20.39088 & -71.56520 & 1.34 &	0.88 & 5.1 &	S & 3-3 & 8-8 & 0 \\
&\\
65665 & IC5054 & 20.89587 & -71.02410 & 1.32 &	0.62 & 1.1 &	? & ? & - & - \\
65794 & ESO286-18 & 20.96403 & -43.37390 & 1.42 & 0.79 & 3.8 &	U & 6-2 & 9-3 & 0.5 \\
65915 & IC5071 & 21.02221 & -72.64490 & 1.53 &	0.67 & 4.8 &	N & 6-6 & 39-23 & 0 \\
66064 & ESO235-53 & 21.08624 & -47.78900 & 1.39 & 0.67 & 3.0 &	N & 9-9 & 22-22 & 0 \\
&\\
66101 & ESO235-57 & 21.10602 & -48.16920 & 1.37 & 0.69 & 3.9 &	- & 0-0 & - & - \\
66530 & IC5096 & 21.30611 & -63.76130 & 1.50 &	0.71 & 4.0 &	N & 8-4 & 18-11 & 0.33 \\
66545 & ESO145-4 & 21.31419 & -57.64030 & 1.34 & 0.61 & 5.0 &	? & b & - & - \\
66617 & ESO287-9 & 21.35448 & -46.15240 & 1.25 & 0.73 & 4.3 &	- & 0-0 & - & - \\
&\\
66836 & NGC7064 & 21.48398 & -52.76610 & 1.52 &	0.76 & 5.3 &	- & 0-0 & - & - \\
67045 & NGC7090 & 21.60794 & -54.55740 & 1.89 &	0.77 & 5.1 &	- & 0-0 & - & - \\
67078 & ESO287-43 & 21.63656 & -43.93260 & 1.30 & 0.77 & 6.1 &	- & 0-0 & - & - \\
67158 & ESO531-22 & 21.67475 & -26.52590 & 1.44 & 1.02 & 4.4 &	S & ?-4 & --11 & - \\
&\\
67782 & ESO288-25 & 21.98822 & -43.86700 & 1.40 & 0.90 & 4.1 &	- & 0-0 & - & - \\
67904 & NGC7184 & 22.04401 & -20.81320 & 1.78 &	0.65 & 4.5 &	N & 3-3 & 9-9 & 0 \\
68223 & IC5171 & 22.18238 & -46.08210 & 1.38 &	0.61 & 3.8 &	N & 6-6 & 11-11 & 0 \\
68329 & NGC7232A & 22.22810 & -45.89360 & 1.33 & 0.70 & 2.0 &	N & 6-6 & 14-14 & 0 \\
&\\
68389 & IC5176 & 22.24848 & -66.84810 & 1.64 &	0.82 & 4.3 &	N & 3-3 & 6-6 & 0 \\
69011 & IC5224 & 22.50824 & -45.99330 & 1.19 &	0.61 & 2.2 &	- & 0-0 & - & - \\
69161 & NGC7307 & 22.56463 & -40.93330 & 1.55 &	0.61 & 5.9 &	N & 11-10 & 31-22 & 0.05 \\
69539 & NGC7361 & 22.70498 & -30.05800 & 1.60 &	0.65 & 4.6 &	S & 5-5 & 14-14 & 0 \\
&\\
69620 & IC5244 & 22.73710 & -64.04230 & 1.46 &	0.73 & 3.0 &	N & 3-2 & 8-4 & 0.2 \\
69661 & NGC7368 & 22.75876 & -39.34150 & 1.48 &	0.75 & 3.1 &	N & 9-6 & 18-14 & 0.2 \\
69707 & IC5249 & 22.78511 & -64.83150 & 1.59 &	1.07 & 6.8 &	- & 0-0 & - & - \\
69967 & NGC7400 & 22.90582 & -45.34670 & 1.41 &	0.63 & 4.0 &	- & 0-0 & - & - \\
\hline
\end{tabular}
%\label{tbl:spiral}
\end{table*}
\newpage
\begin{table*}[ht]
\caption{ (Table 1, continuation)}
\begin{tabular}{ccccccccccc}
\hline
pgc & galaxy-name & al2000 & de2000 & logd25 &	logr25 & t &
\emph{N/S} & \emph{(wa)E-W} ($^o$) & $\beta$ ($^o$) & $\alpha_{s}$ \\
\hline
70025 & NGC7416 & 22.92829 & -5.49650 & 1.50 &	0.68 & 3.0 &	S & 4-4 & 9-9 & 0 \\
70070 & IC5269B & 22.94356 & -36.24970 & 1.58 &	0.72 & 5.6 &	- & 0-0 & - & - \\
70081 & IC5264 & 22.94796 & -36.55430 & 1.39 &	0.72 & 2.4 &	N & 5-5 & 16-16 & 0 \\
70084 & MCG-2-58-11 & 22.94754 & -8.96760 & 1.31 & 0.67 & 4.7 &	S & 7-4 & 16-8 & 0.27 \\
&\\
70142 & IC5266 & 22.97244 & -65.12970 & 1.24 &	0.61 & 3.1 &	- & 0-0 & - & - \\
70324 & NGC7462 & 23.04623 & -40.83400 & 1.62 &	0.71 & 4.1 &	- & 0-0 & - & - \\
71309 & ESO291-24 & 23.39472 & -42.40210 & 1.22 & 0.61 & 5.0 &	? & ? & - & - \\
71800 & IC5333 & 23.58140 & -65.39590 & 1.24 &	0.71 & 3.4 &	LN & 7-0 & 22-- & 1 \\
&\\
71948 & ESO240-11 & 23.63039 & -47.72630 & 1.74 & 0.92 & 4.8 &	- & 0-0 & - & - \\
72178 & ESO292-14 & 23.70990 & -44.90460 & 1.43 & 0.88 & 6.5 &	N & 2-5 & 6-11 & 0.43 \\
\hline
\end{tabular}
%\label{tbl:spiral}
\end{table*}

\begin{table*}[ht]
\caption{Sample of Lenticular Galaxies}
\begin{tabular}{cccccccc}
\hline
pgc & galaxy-name & al2000 & de2000 & logd25 &	logr25 & t &	\emph{N/S}  \\
\hline
5210 & NGC530 & 1.41159 & -1.58770	       &  1.24 &  0.58 & -0.3 &  -   \\
5430 & NGC560 & 1.45706 & -1.91310	     &    1.30 & 0.57 & -2.4 &  -   \\
6117 & NGC643 & 1.65349  &-75.01100	       &  1.22 & 0.60 & -0.2 &  -   \\
12662 & ESO301-9 & 3.38191 & -42.18790 &	1.30 & 0.69 & -1.7 &  -   \\
&\\
13169 & NGC1355 & 3.55654 & -4.99880	 &        1.20 & 0.61 & -2.1 &  -   \\
13241 & ESO548-47 & 3.57875 & -19.02900 &	1.40 & 0.60 & -0.8 &  -   \\
13277 & IC335 & 3.59187 & -34.44660	        & 1.37 & 0.58 & -1.6 &  -  \\
14495 & NGC1529  & 4.12195  & -62.89900	 &        1.10 & 0.58 & -2.2  &  - \\
&\\
15388 & IC2085  & 4.52344 & -54.41690	     &    1.35 & 0.64 & -1.2 &  -  \\
19811 & NGC2310 & 6.89821 & -40.86220	    &     1.62 & 0.74 & -1.9 &  -  \\
24195 & ESO562-23 & 8.60971 & -20.47010 &	1.36  & 0.61 & -0.9 &  -  \\
24966 & ESO371-26 & 8.90906 & -32.93740 &	1.49  & 0.68 & -1.5 &  -  \\
&\\
25202 & ESO90-12 & 8.97290 & -66.72830	 &1.34 & 0.62 & -1.8 &  -  \\
25943 & ESO433-8 & 9.20360 & -30.91120	 & 1.32 & 0.66 & -1.9 &  -  \\
30177 & NGC3203 & 10.32623 & -26.69820	 & 1.45 & 0.65 & -1.3 &  -  \\
30792 & NGC3250D & 10.46610 & -39.81490 &	1.24 & 0.71 & -1.9 &  -  \\
&\\
30938 & IC2584  &10.49771  & -34.91160	 &        1.16 & 0.57 & -2.0 &  -  \\
31369 & MCG-2-27-9  &10.59093 & -14.12990 &	1.32 & 0.58 & -1.0 &  -  \\
31504 & ESO437-15 & 10.61611 & -28.17810	 & 1.33 & 0.58 & -2.0 &  -  \\
36417 & NGC3831 & 11.72187 & -12.87700	 & 1.38 & 0.60 & -0.8 &  -  \\
&\\
37326 & NGC3957  & 11.90028 & -19.56820 &	1.49 & 0.64  &-1.0 &  -  \\
42486 & NGC4603C & 12.67865 & -40.76350 &	1.24 & 0.59 & -2.0 &  -  \\
43929 & NGC4784 & 12.91030 & -10.61300	 & 1.19  & 0.60 & -1.8 &  -  \\
45650 & MCG-3-34-4 & 13.16221 & -16.60210 &	1.35 & 0.59 & -1.0 &  -  \\
&\\
46081 & NGC5038 & 13.25063 & -15.95170	 & 1.19 & 0.62 & -1.9 &  -  \\
46150 & NGC5047 & 13.26347 & -16.51940	 & 1.43 & 0.71 & -2.0 &  -  \\
46166 & NGC5049 & 13.26648 & -16.39550	 & 1.28 & 0.64 & -2.0 &  -  \\
46525 & NGC5084 & 13.33799 & -21.82700	 & 2.03 & 0.60 & -1.8 &  -  \\
&\\
49006 & ESO445-42 & 13.81359  &-31.15510	 & 1.14 & 0.74   & -0.4 &  -  \\
49300 & ESO445-65 & 13.87963 & -29.92950	 & 1.18 & 0.62  & -2.2 &  -  \\
49840 & ESO384-26 & 14.00407 & -34.03760 &	1.19 & 0.57 & -1.9 &  -  \\
50242 & IC4333 & 14.08889 & -84.27290	 &        1.20 & 0.62  & -1.8 &  -  \\
&\\
62692 & NGC6725 & 19.03230  & -53.86470 &	1.36 & 0.65 & -2.0 &  -  \\
63039 & ESO184-53 & 19.30533 & -53.47730	 & 1.10 & 0.60 & -1.8 &  -  \\
63049 & NGC6771 & 19.31105  & -60.54560 &	1.37 & 0.64  & -1.0 &  -  \\
65055 & ESO234-53 & 20.60682 & -49.25780 &	1.28 & 0.59  & -2.0 &  -  \\
&\\
66908 & ESO47-34 & 21.52875 & -76.48040 &	1.21 & 0.66 & -2.0 &  -  \\
69638 & NGC7359 & 22.74651 & -23.68700	 & 1.35 & 0.58 &  -1.8 &  -  \\
\hline
\end{tabular}
\label{tbl:lenti}
\end{table*}


\begin{thebibliography}{}
  
   \bibitem[1990]{battaner} Battaner, E., Florido, E., \&
   S\'anchez-Saavedra, M.L. 1990, A\&A, 236, 1

   \bibitem[1991]{battaner2} Battaner, E., Garrido, J.L.,
   S\'anchez-Saavedra, M.L., \&Florido, E., 1991, A\&A, 251, 402

   \bibitem[1992]{binney} Binney, J.J. 1992, ARAA, 30, 51
   
   \bibitem[1981]{bosma} Bosma, A. 1981,
      AJ, 86, 1791

   \bibitem[2002]{nieves} Castro-Rodr\'{\i}guez, N.,
   L\'opez-Corredoira, M., S\'anchez-Saavedra, M.L., \& Battaner,
   E. 2002, A\&A, in press

   \bibitem[1995]{dubinski} Dubinski, J., \& Kuijken, K. 1995,
      ApJ, 442, 492

   \bibitem[1991]{florido} Florido, E., Prieto, M., Battaner, E.,
      Mediavilla, E., \& S\'anchez-Saavedra, M.L. 1991,
      A\&A, 242, 301
  
   \bibitem[2001]{garciaruiz01a} Garc\'{\i}a-Ru\'{\i}z, I. 2001,
      Ph. D. Thesis,
      Groningen University

   \bibitem[2000]{garciaruiz01b} Garc\'{\i}a-Ru\'{\i}z, I., Kuijken,
       K., \& Dubinski, J. 2000, MNRAS, submitted. Preprint
       astro-ph/0002057  

   \bibitem[1997]{grijs} de Grijs, R. 1997,
      Ph. D. Thesis,
      Groningen University

   \bibitem[1998]{jimenez98a} Jim\'enez-Vicente, J. 1998,
      Ph. D. Thesis,
      Granada University

   \bibitem[1998]{jimenez98b} Jim\'enez-Vicente, J., Porcel, C.,
      S\'anchez-Saavedra, M.L., \& Battaner, E. 1998,
      Ap\&SS, 253, 225

   \bibitem[1959]{kant} Kahn, F.D., \& Woltjer, L. 1959,
      AJ, 130, 705
   \bibitem[2001]{kuijken} Kuijken, K., \& Garc\'{\i}a-Ru\'{\i}z,
      I. 2001, \emph{Galaxy Disks and Disk Galaxies, Vol. 230},
      ed. J.G. Funes, S.J. and E.M. Corsini, p. 401, ASP Conference
      Serie 

   \bibitem[2002]{lopez-corredoira} L\'opez-Corredoira, M., Betancort,
      J.A., \& Beckman, J. 2002,
      A\&A, to be published

    \bibitem[1989]{ostriker} Ostriker, E.C., \&  Binney, J.J. 1989,
       MNRAS, 237, 785

   \bibitem[1998]{reshetnikov98} Reshetnikov, V., \& Combes, F. 1998,
     A\&A, 337, 9

   \bibitem[1999]{reshetnikov99} Reshetnikov, V., \& Combes, F. 1999,
     A\&A, 138, 101R

   \bibitem[2002]{reshetnikov2002} Reshetnikov, V., Battaner, E.,
   Combes, F., \& Jim\'enez-Vicente, J. 2002, A\&A, 382, 513

   \bibitem[1990]{sanchez-saavedra}  S\'anchez-Saavedra, M.L.,
      Battaner, E., \& Florido, E. 1990, 
      MNRAS, 246, 458   
    
   \bibitem[1976]{sancisi} Sancisi, R. 1976,
      A\&A, 53, 159

   \bibitem[2001]{schwarzkopf} Schwarzkopf, U., \& Dettmar, R.J. 2001,
      A\&A, 373, 402

   \bibitem[1988]{sparke} Sparke, L., \& Casertano, S. 1988,
      MNRAS, 234, 873

   \bibitem[1998]{weinberg} Weinberg, M.D. 1998,
      MNRAS, 299, 499

  
\end{thebibliography}
\end{document}